\documentclass[twocolumn]{aastex62}
\usepackage{natbib}
\usepackage{amsmath}

\newcommand{\logg}{\ensuremath{\log g}}

\newcommand{\msun}{$M_{\odot}$}
\newcommand{\kms}{${\rm km\;s^{-1}}$}
\usepackage{graphicx}

\received{Nov 05, 2019}
\revised{Nov XX, 2019}
\accepted{Dec XX, 2019}
\submitjournal{ApJ}

\shorttitle{Detailed chemical analysis of J0815+4729}
\shortauthors{Gonz\'alez Hern\'andez et al.}

\begin{document}

\title{The extreme CNO-enhanced composition of the primitive iron-poor dwarf 
star J0815+4729\footnote{Based on observations made with Keck I telescope at 
Mauna Kea Observatory, Hawaii, USA}}

\correspondingauthor{Jonay~I. Gonz\'alez Hern\'andez}
\email{jonay@iac.es}

\author[0000-0002-0264-7356]{Jonay~I. Gonz\'alez Hern\'andez}
\affiliation{Instituto de Astrof\'{\i}sica de Canarias, 
V\'{\i}a L\'actea, 38205 La Laguna, Tenerife, Spain\\}
\affiliation{Universidad de La Laguna, Departamento de Astrof\'{\i}sica, 
38206 La Laguna, Tenerife, Spain\\} 

\author[0000-0001-5200-3973]{David~S. Aguado}
\affiliation{Institute of Astronomy, University of Cambridge, Madingley Road, 
Cambridge CB3 0HA, UK \\}             
            
\author[0000-0002-0084-572X]{Carlos Allende Prieto}
\affiliation{Instituto de Astrof\'{\i}sica de Canarias, 
V\'{\i}a L\'actea, 38205 La Laguna, Tenerife, Spain\\}
\affiliation{Universidad de La Laguna, Departamento de Astrof\'{\i}sica, 
38206 La Laguna, Tenerife, Spain\\} 
             
\author[0000-0002-6523-9536]{Adam~J. Burgasser}
\affiliation{Center for Astrophysics and Space Science, University of California 
San Diego, La Jolla, CA 92093, USA\\}
            
\author[0000-0003-3767-7085]{Rafael Rebolo}
\affiliation{Instituto de Astrof\'{\i}sica de Canarias, 
V\'{\i}a L\'actea, 38205 La Laguna, Tenerife, Spain\\}
\affiliation{Universidad de La Laguna, Departamento de Astrof\'{\i}sica, 
38206 La Laguna, Tenerife, Spain\\} 
\affiliation{Consejo Superior de Investigaciones Cient\'{\i}ficas, 
28006 Madrid, Spain\\}



\begin{abstract}
We present an analysis of high-resolution Keck/HIRES spectroscopic observations of 
J0815+4729, an extremely carbon-enhanced, iron-poor dwarf star. 
These high-quality data allow us to derive a metallicity of [Fe/H]$=-5.49{\pm}0.14$ 
from the three strongest \ion{Fe}{1} lines and to measure
a high [Ca/Fe]~$=0.75{\pm}0.14$.
The large carbon abundance of A(C)~$=7.43{\pm}0.17$ (or 
[C/Fe]~$\sim 4.49{\pm}0.11$) places this star in the upper 
boundary of the low-carbon band in the A(C)-[Fe/H] diagram, suggesting no contamination 
from a binary AGB companion.  
We detect the oxygen triplet at 777nm for the first time in an ultra-metal poor star, 
indicating a large oxygen-to-iron abundance ratio of [O/Fe]~$=4.03{\pm}0.12$ 
(A(O)~$=7.23{\pm}0.14$), 
significantly higher than the previously most metal-poor dwarf J2209-0028 with an 
oxygen triplet detection with [O/Fe]~$\sim2.2$~dex at [Fe/H]~$\sim -3.9$. 
Nitrogen is also dramatically 
enhanced with (A(N)~$=6.75{\pm}0.08$) and an abundance ratio 
[N/Fe]~$\sim 4.41{\pm}0.08$. 
We also detect Ca, Na and Mg, while provide upper limits for eight other elements.
The abundance pattern of J0815+4729 resembles that of HE~1327-2326, indicating that 
both are second-generation stars contaminated by a $\sim 21-27$~\msun~single, 
zero-metallicity low-energy supernova with very little mixing and substantial fallback.
The absence of lithium implies an upper-limit abundance A(Li)~$<1.3$~dex, about
0.7~dex below the detected Li abundance in J0023+0307 which has a similar
metallicity, exacerbating the cosmological lithium problem.
\end{abstract}



\section{Introduction} \label{sec:intro}

Some of the most metal-poor low-mass stars formed in the very Early Universe
are still observable in the Galactic Halo. These stars are expected to 
originate from a mixture of primordial material from the Big Bang and matter ejected 
from the few first supernovae.
The chemical pattern of these stars holds crucial information about the early epochs of 
the Universe, such as the properties of the first stars (so-called Population III stars) 
and first supernovae, the early chemical evolution of the Universe, and the formation 
of low-mass stars in the Early Universe.

Over the last decades, a significant observational effort has been focused 
on the detection and characterization of the most metal-poor stars. 
These are initially identified in large spectroscopic surveys such as 
Hamburg/ESO~\citep[HE;][]{chr01aa} or
Sloan Digital Sky Survey~\citep[SDSS;][]{yor00aj}, or in
narrow-filter photometric surveys such as Skymapper~\citep{kell07pasa}
or Pristine~\citep{sta17mnras}. 
Follow-up high quality medium- and high-resolution spectra are used to confirm the 
metal-poor nature of these stars and provide their detailed chemical abundances.
To date, only 7 stars are known to have [Fe/H]~$< -5$  
\citep{chr04apj,fre08apj,kel14nat,bon15aa,agu18apjlI,agu18apjlII,nord19mnras} 
and all are carbon-enhanced (CEMP) with 
carbon abundances in the range A(C)$=6-7.5$~dex.

The metal-poor halo star SDSS~J081554.26+472947.5 was first identified in the SDSS/BOSS 
spectroscopic database and confirmed as an extremely carbon-rich, iron-poor star 
by~\citet{agu18apjlI} using follow-up medium resolution observations obtained with 
OSIRIS at the 10.4m GTC telescope. 
Here we present new high-resolution high-quality HIRES spectroscopic observations 
revealing the unique abundance pattern of this star.

\begin{figure*}
\begin{center}
{\includegraphics[clip=true,width=85mm,angle=0]{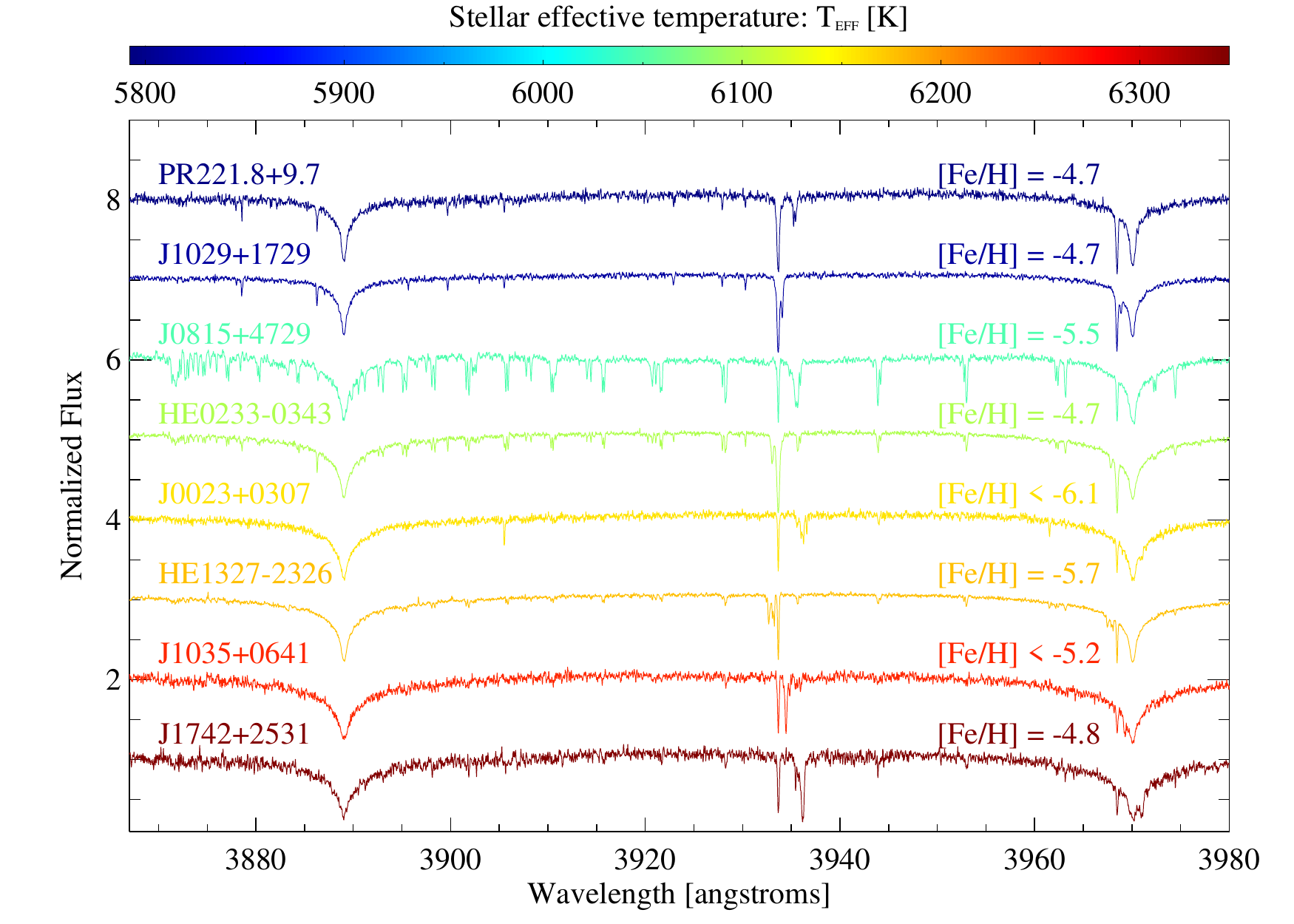}}
{\includegraphics[clip=true,width=85mm,angle=0]{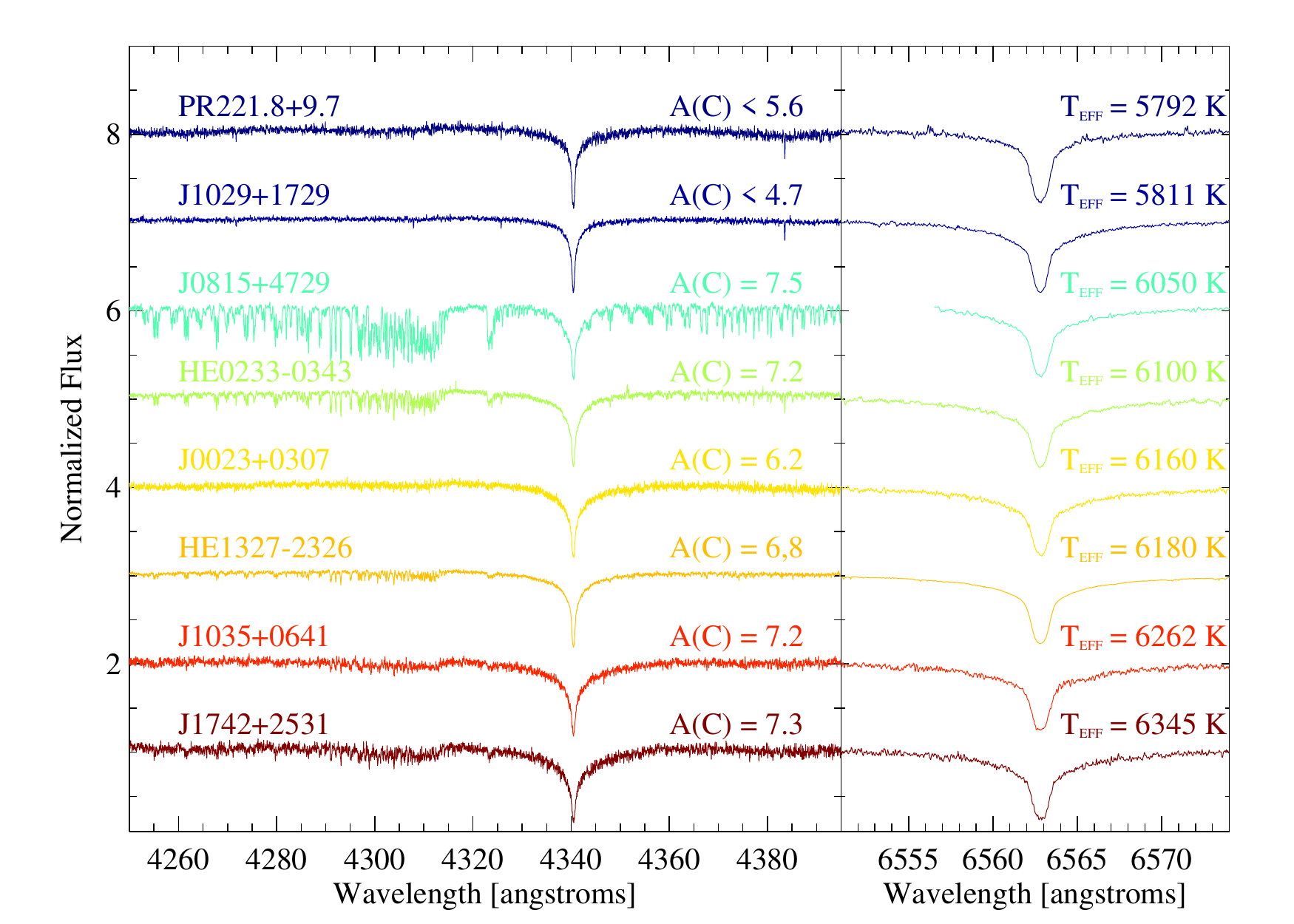}}
\end{center}
\caption{
Keck/HIRES high-resolution spectra of the star J0815+4729 in three different spectral 
regions, compared to VLT/UVES high-resolution spectra of other unevolved, extremely 
iron-poor stars
\citep{bon18aa,fre08apj,han14apj,agu19apjl,sta18mnras,caf12aa}. 
The stars are sorted and colored by their effective temperature from top to bottom.
\label{fig:spe}
}
\end{figure*}

\section{Observations} \label{sec:obs}

High resolution spectroscopic observations of J0815+4729 were carried out on 
2018 December 10 (UT) using the High Resolution Echelle 
Spectrometer~\citep[HIRES,][]{vog94spie} 
on the Keck-I telescope at the Mauna Kea observatory. 
We used the cross-disperser HIRESr 
with a slit width of 1.15\arcsec and 2x2 binning, providing spectra over the range 
$\lambda\lambda 340-780$~nm at a resolving power of $R\sim 37,500$. 
Seven spectra of 2400s were obtained over a seeing 
of 1.0--1.3\arcsec and airmasses of 1.1--1.3.
The individual spectra had S/N of 15, 21, 31, 40, 42 at $\lambda =$ 395, 440, 518, 
670 and 775 nm, respectively.
We used the {\sc Makee} software to 
reduce the data, and the IRAF package {\sc ECHELLE} to calibrate the individual 
spectra and normalize each order using a third-order polynomial continuum correction. 

Radial velocities (RV) were derived by cross-correlating each barycenter-corrected 
observed spectrum with a synthetic spectrum in the range $\lambda\lambda 418-432$~nm. 
This region encompasses many individual lines of the CH A-X band that are pronounced 
in this carbon-enriched star (see Fig.~\ref{fig:spe}).
We measure a mean radial velocity of $v_{\rm R}=-123.1 \pm 0.4$~\kms,  
consistent with the $v_{\rm R}=-95\pm 23$~\kms\ reported in the discovery 
paper from GTC/OSIRIS spectrum~\citep{agu18apjlI}, with  
$v_{\rm R}=-118\pm 5$~\kms\ determined from the discovery SDSS/BOSS 
spectrum~\citep{daw13aj}, and with $v_{\rm R}\sim -124.4$~\kms\ from a LAMOST 
spectrum~\citep{li18apjs}. 
We find no evidence of RV variability from these measurements.

The HIRES spectra were RV corrected, co-added and rebinned using wavelength 
steps of 0.030, 0.042, and 0.055~{\AA}~pixel$^{-1}$, 
in the blue ($\lambda\lambda 340-473$~nm), green ($\lambda\lambda 478-631$~nm) and 
red ($\lambda\lambda 632-782$~nm) detector regions, respectively. 
In Fig.~\ref{fig:spe} we compare the HIRES 
spectrum of J0815+4729 to VLT/UVES spectra of other extremely 
iron-poor stars available in the 
ESO\footnote{ESO program IDs: 076.D-0546(A), 075.D-0048(A), 
077.D-0035(A), 091.D-0288(A), 096.D-0468(A), 0101.D-0149(A), 
189.D-0165(F), 286.D-5045(A), 299.D-5042(A)} 
Archive Science 
Portal\footnote{Available at \url{http://archive.eso.org/wdb/wdb/eso/uves/form}.} 
For proper comparison, we have processed the UVES data similarly: 
individual spectra have been normalized using a fifth-order 
polynomial continuum correction based on our own automated IDL-based routine, 
corrected for barycentric motion and star's radial velocity,
and co-added and rebinned to the same wavelength step as the final HIRES spectrum.

\begin{figure*}
\begin{center}
{\includegraphics[clip=true,width=85mm,angle=0]{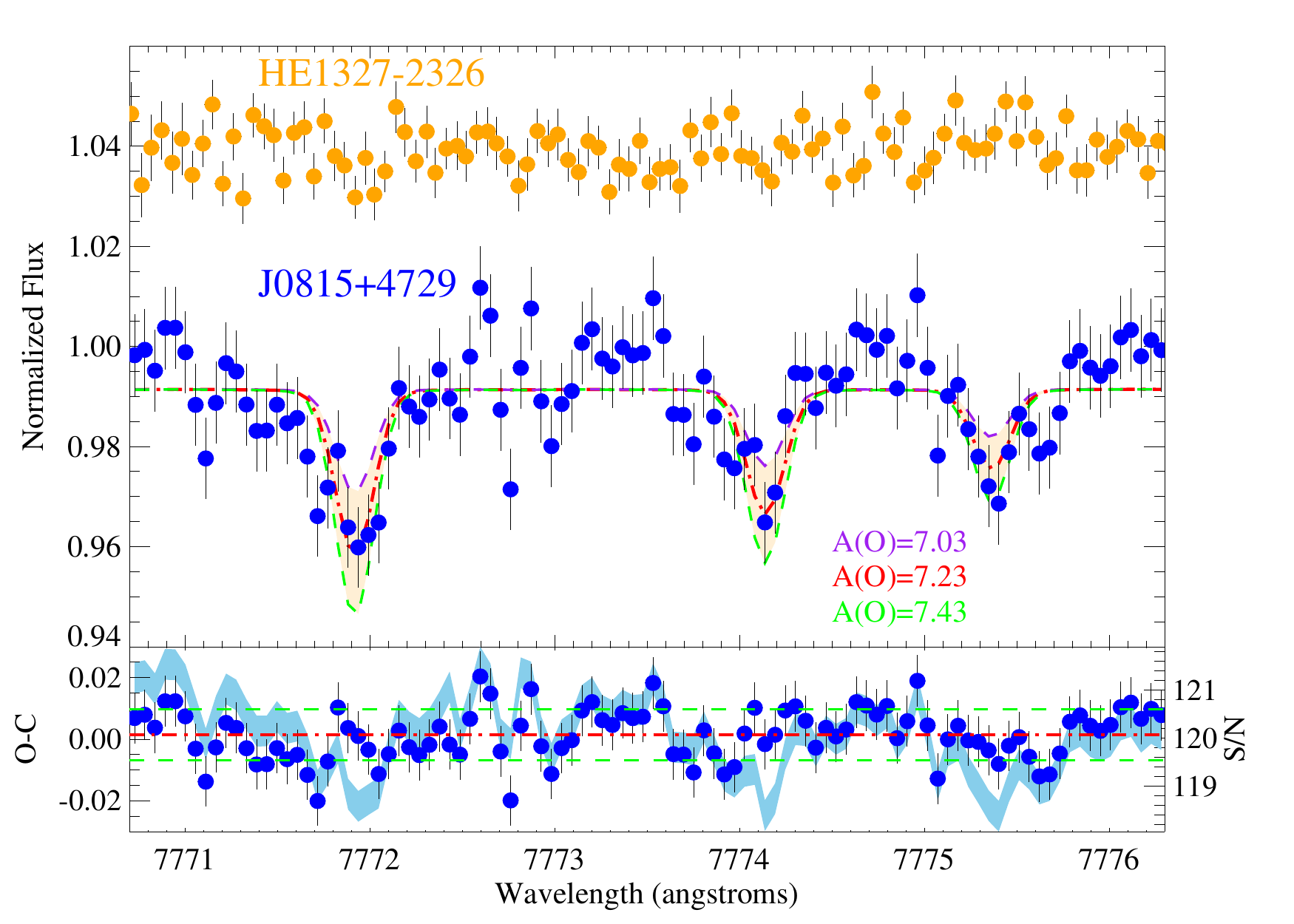}}
{\includegraphics[clip=true,width=85mm,angle=0]{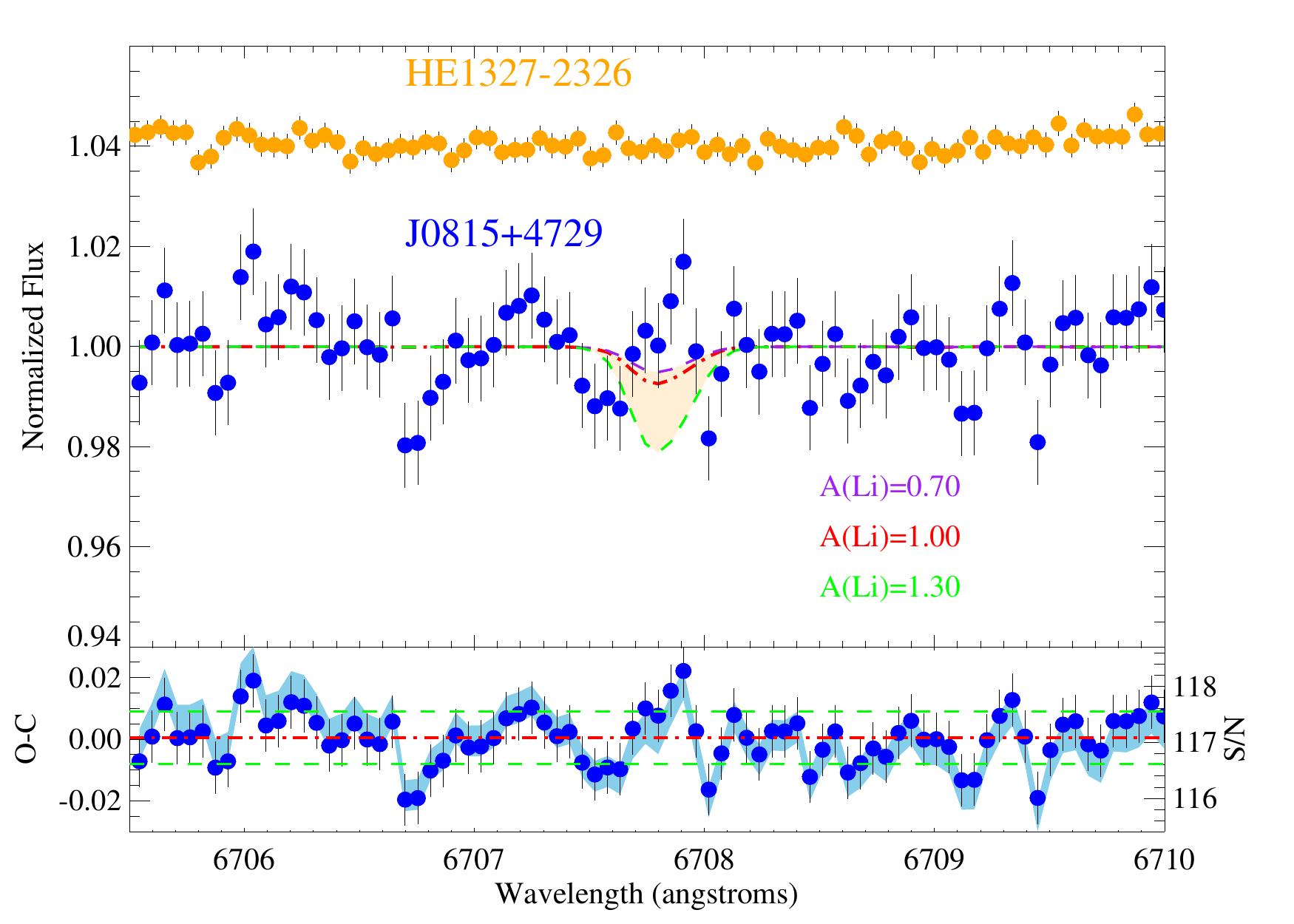}}
{\includegraphics[clip=true,width=85mm,angle=0]{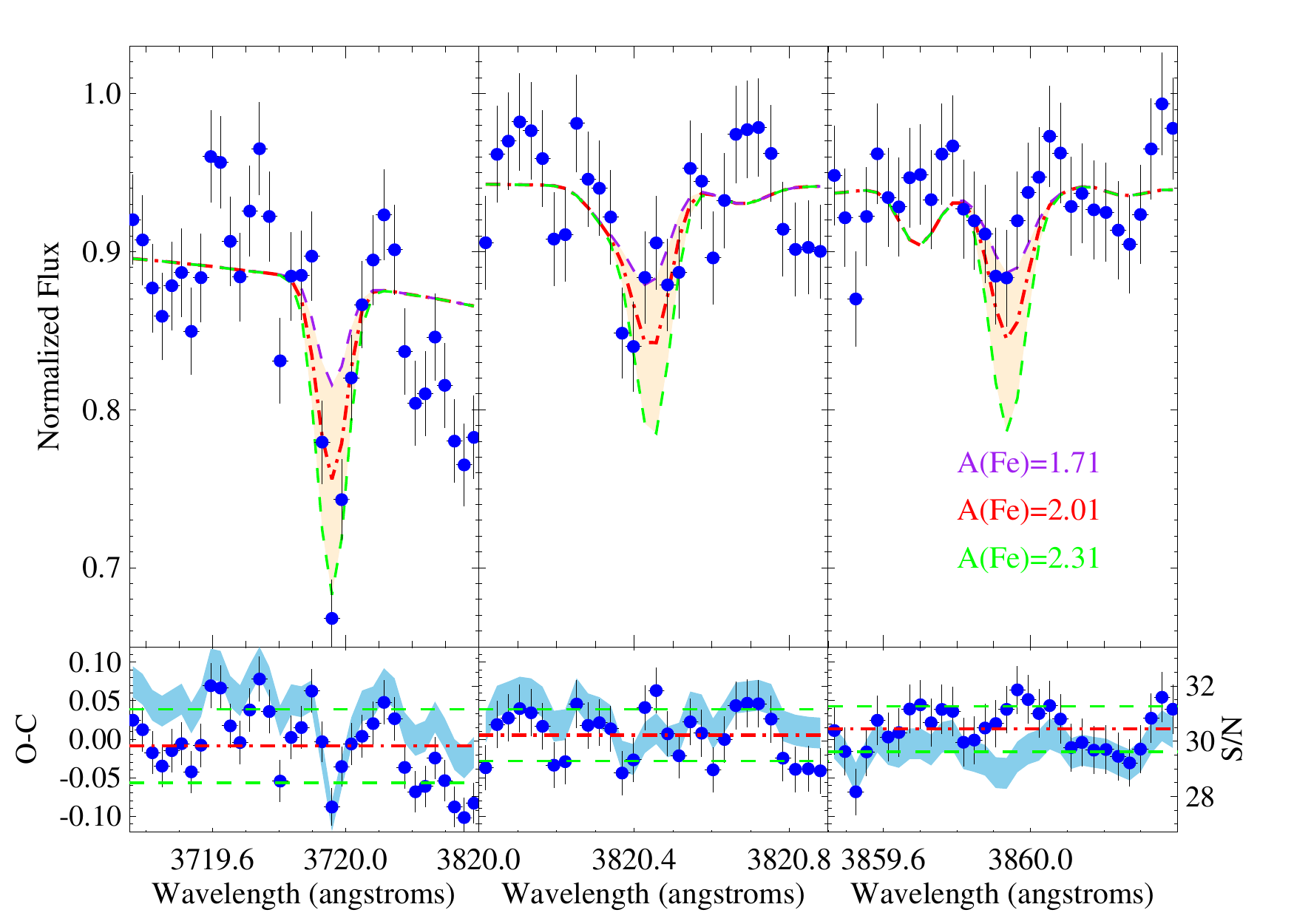}}
{\includegraphics[clip=true,width=85mm,angle=0]{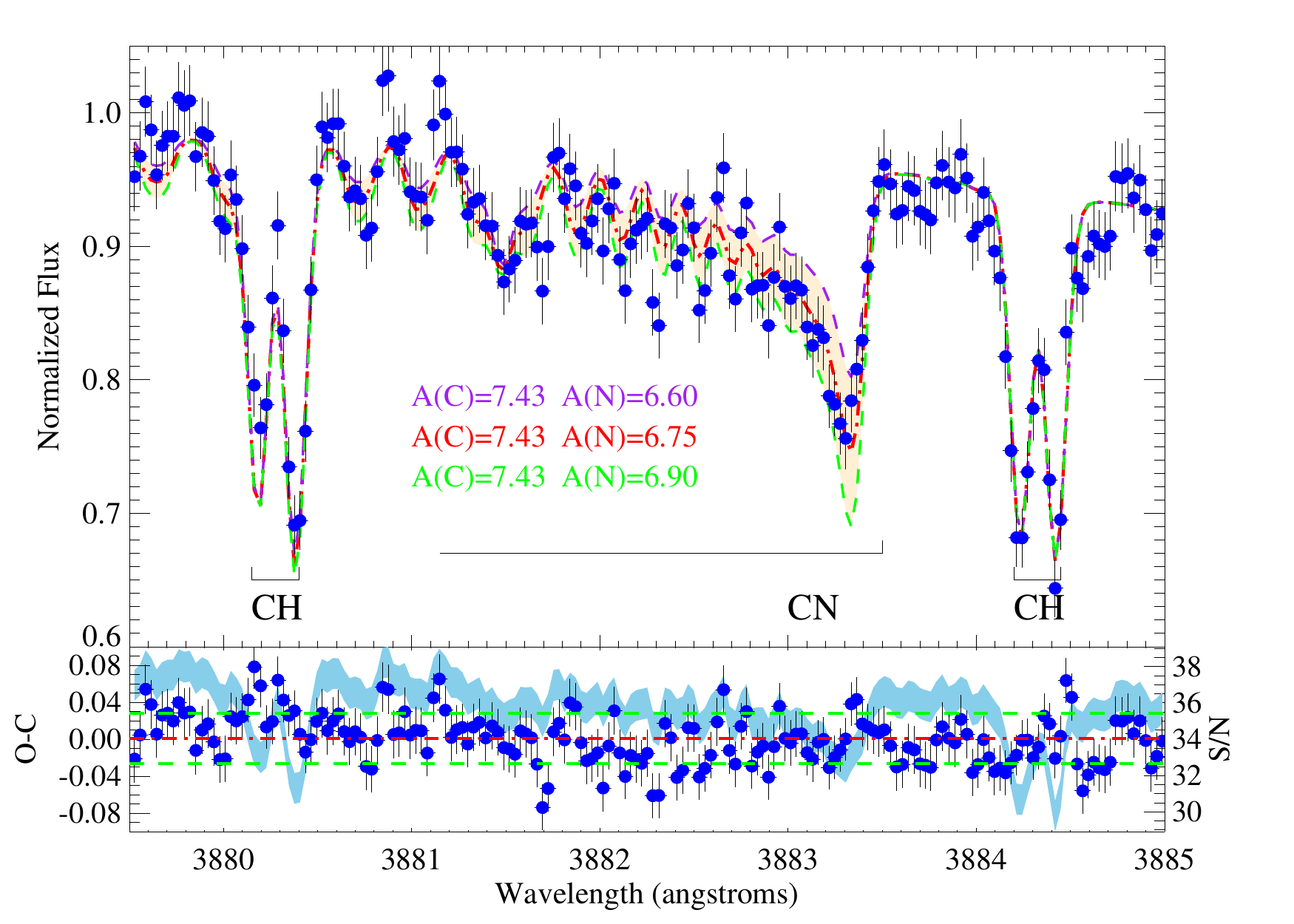}}
{\includegraphics[clip=true,width=85mm,angle=0]{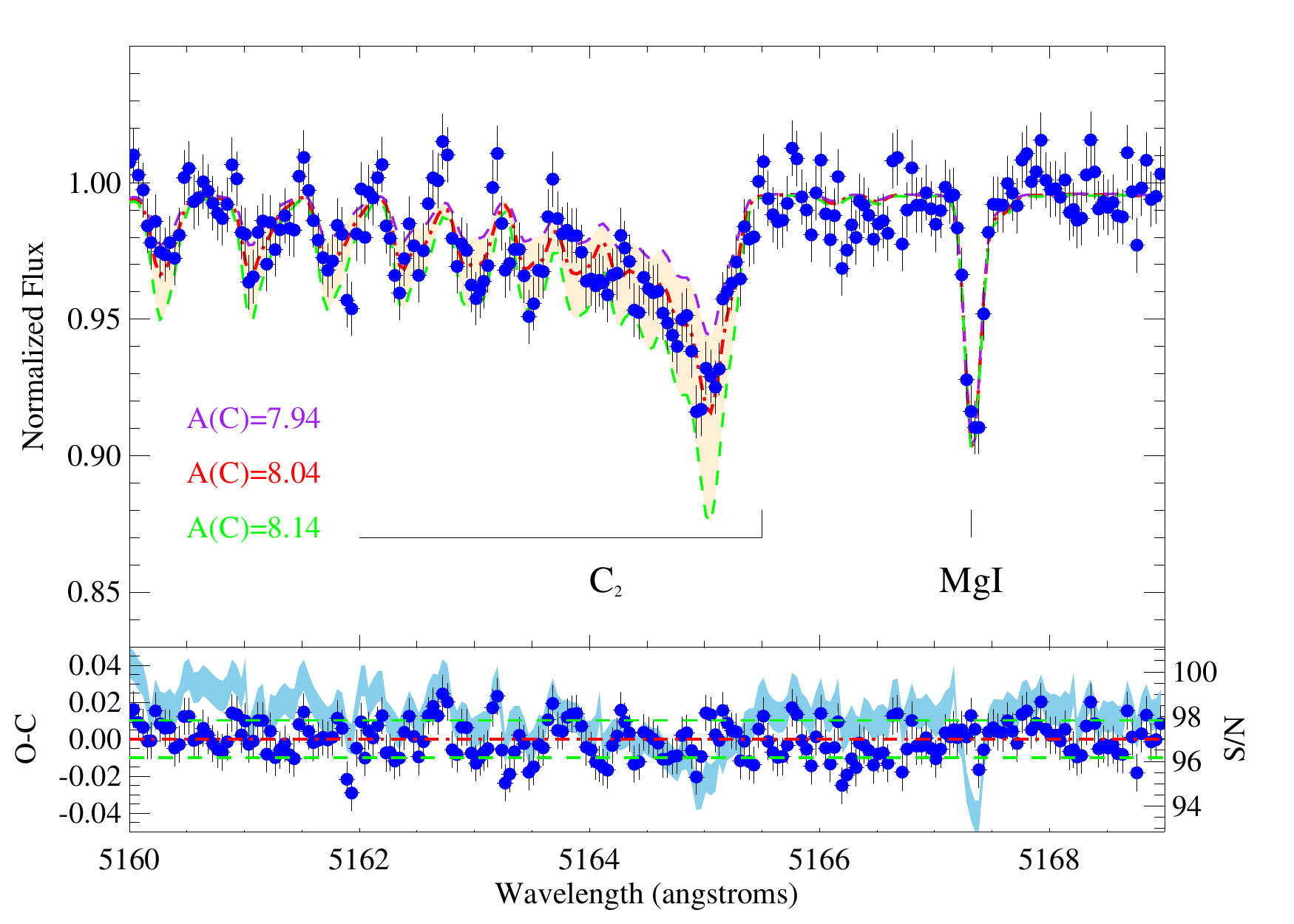}}
\end{center}
\caption{
Specific spectral features in the Keck/HIRES spectrum of J0815+4729 (blue points 
with error bars), compared to  
{\sc SYNPLE} synthetic spectra (the best-fit abundance is shown as red 
dashed-dotted line). Shown are lines of oxygen (top left), lithium 
non-detection (top right), 
iron (middle left), nitrogen (middle right), carbon C$_2$ (bottom). 
Top panels also show for comparison the UVES spectrum of HE 1327-2326~\citep{fre08apj}, 
rebinned to the same wavelength step as the HIRES spectrum of J0815+4729. 
The bottom frame of each panel provides the residuals of the observations minus 
the best-fit model (blue points with error bars) and the uncertainty range 
(green dashed lines), and the signal-to-noise as the light blue curve.
\label{fig:oxy}
}
\end{figure*}

\section{Stellar parameters} \label{sec:ste}

\begin{table}[!ht]
\renewcommand{\thetable}{\arabic{table}}
\centering
\caption{Element 1D LTE abundances of J0815+4729\label{tab:abu}}
\begin{tabular}{lrrrrr}
\hline\hline
Species & ${\rm A}_\odot({\rm X})$ & ${\rm A}({\rm X})$ & 
$\delta {\rm A}({\rm X})$ & \mbox{[X/H]} & $N$ \\
\hline
\decimals
\ion{Li}{1} & 1.05 &  $<$ 1.30  &  --   &   --      & 6708 \\
CH          & 8.43 &      7.43  &  0.17 &  $-$1.00  & --  \\
C$_2$       & 8.43 &      8.04  &  0.08 &  $-$0.39  & --  \\
CN          & 7.83 &      6.75  &  0.08 &  $-$1.08  & --  \\
\ion{O}{1}  & 8.69 &      7.23  &  0.14 &  $-$1.46  & 3 \\
\ion{Na}{1} & 6.24 &      3.68  &  0.13 &  $-$2.56  & 2 \\
\ion{Mg}{1} & 7.60 &      3.77  &  0.08 &  $-$3.83  & 6 \\
\ion{Al}{1} & 6.45 &  $<$ 1.50  &  --   & $<-$4.95  & 3961 \\
\ion{Si}{1} & 7.51 &  $<$ 3.30  &  --   & $<-$4.21  & 3905 \\
\ion{Ca}{2} & 6.34 &      1.60  &  0.18 &  $-$4.74  & 2 \\         
\ion{Ti}{2} & 4.95 &  $<$ 0.70  &  --   & $<-$4.25  & 3760 \\
\ion{Cr}{1} & 5.64 &  $<$ 1.50  &       & $<-$4.14  & 4254 \\
\ion{Fe}{1} & 7.50 &      2.01  &  0.14 &  $-$5.49  & 3 \\
\ion{Ni}{1} & 6.22 &  $<$ 1.90  &  --   & $<-$4.32  & 3858 \\
\ion{Sr}{2} & 2.87 & $<-$ 1.60  &  --   & $<-$4.47  & 4077 \\
\ion{Ba}{2} & 2.18 & $<-$ 1.40  &  --   & $<-$3.58  & 4554 \\
\hline
\multicolumn{6}{l}{
$^\dagger$ Solar abundances from \citet{asp09sun}}\\
\multicolumn{6}{l}{
${\rm A}_\odot({\rm X})= \log \epsilon_\odot({\rm X})=
\log[N_\odot({\rm X})/N_\odot({\rm H})]+12$
}\\
\multicolumn{6}{l}{
$[{\rm X}/{\rm H}]={\rm A}({\rm X})-{\rm A}_\odot({\rm X})=
\log \epsilon({\rm X})-\log \epsilon_\odot({\rm X})$
}\\
\multicolumn{6}{l}{
$^{\dagger\dagger}$ Total abundance errors from the uncertainties}\\
\multicolumn{6}{l}{ 
in $\delta T_{\rm eff}=100$~K,  $\delta \logg=0.20$~dex, $\delta \xi =0.5$~\kms,}\\
\multicolumn{6}{l}{ 
and the statistical error}\\
\multicolumn{6}{l}{
$^\ddagger$ Number of spectral features or wavelength}\\
\multicolumn{6}{l}{
in Angstroms if only one feature}
\end{tabular}
\end{table}

Following \citet{agu19apjl} we analysed the spectrum of J0815+4729 using
a grid of synthetic spectral models spanning $-7 \leq {\rm [Fe/H]} \leq -4$, 
$-1 \leq {\rm [C/Fe]} \leq 7$, $4750\,{\rm K} \leq T_{\rm eff} \leq 7000\,{\rm K}$, and
$1 \leq \log (g) \leq 5$, computed with the code ASS$\epsilon$T~\citep{koe08apj} 
and model atmospheres from Kurucz ATLAS 9~\citep{mez12aj}. The micro-turbulence, $\xi$, 
and $\alpha$-element abundance were fixed at 2~\kms~and [$\alpha$/Fe]=0.4, respectively. 
We used the {\sc FERRE} 
code\footnote{Available at \url{https://github.com/callendeprieto/ferre}}
\citep[see e.g.][]{agu17aaI} to fit the whole
HIRES spectrum of J0815+4729 and separately to fit only the Balmer H~I lines. Both 
methods provide the same effective temperature, $T_{\rm eff} \sim 6050$~K, and  
surface gravity, $\logg=4.7$~dex. This $T_{\rm eff}$ is lower than our previous 
estimates from medium resolution OSIRIS, BOSS and ISIS spectra, the former yielding 
$T_{\rm eff}=6215\pm82$~K ~\citep{agu18apjlI}. We also estimate a 
$T_{\rm IRFM} \sim 6196\pm 185$~K by applying the infrared flux 
method~\citep{gon09irfm} to 2MASS $J$ and Johnson $V$ magnitudes (the latter 
transformed from SDSS $g$ and $r$ photometry), and assuming a reddening of 
E(B-V)$=0.073$~\citep{sche98apj}. 
Note that 2MASS $HK_s$ magnitudes were not used due to their large uncertainties.
We also apply the $T_{\rm eff}$--(V-J),[Fe/H]
calibration in \citep{cas10aa} to estimate a $T_{\rm IRFM} \sim 6237\pm 228$~K, 
consistent with the previous values.

\begin{figure}
\begin{center}
{\includegraphics[clip=true,width=85mm,angle=0]{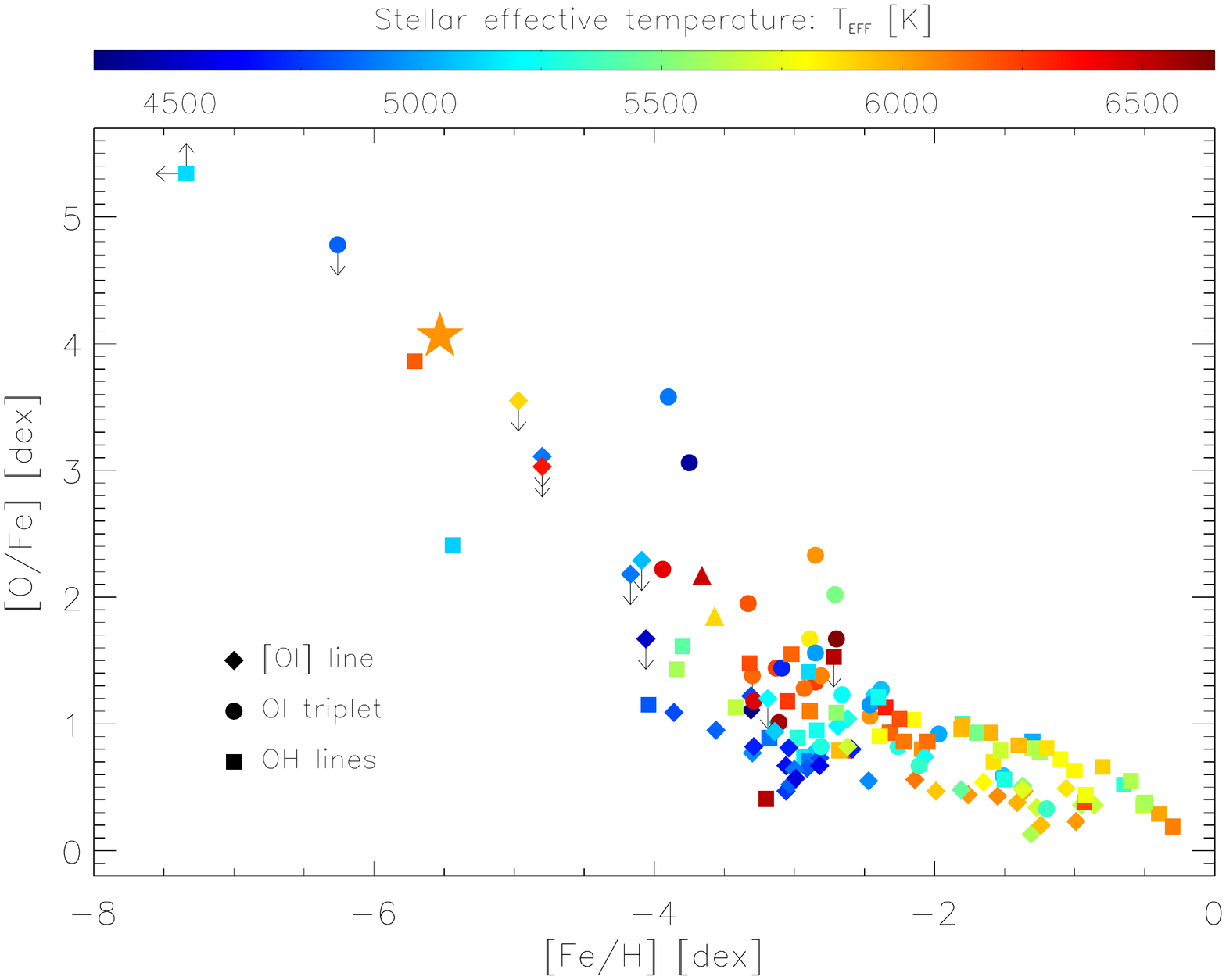}}
{\includegraphics[clip=true,width=85mm,angle=0]{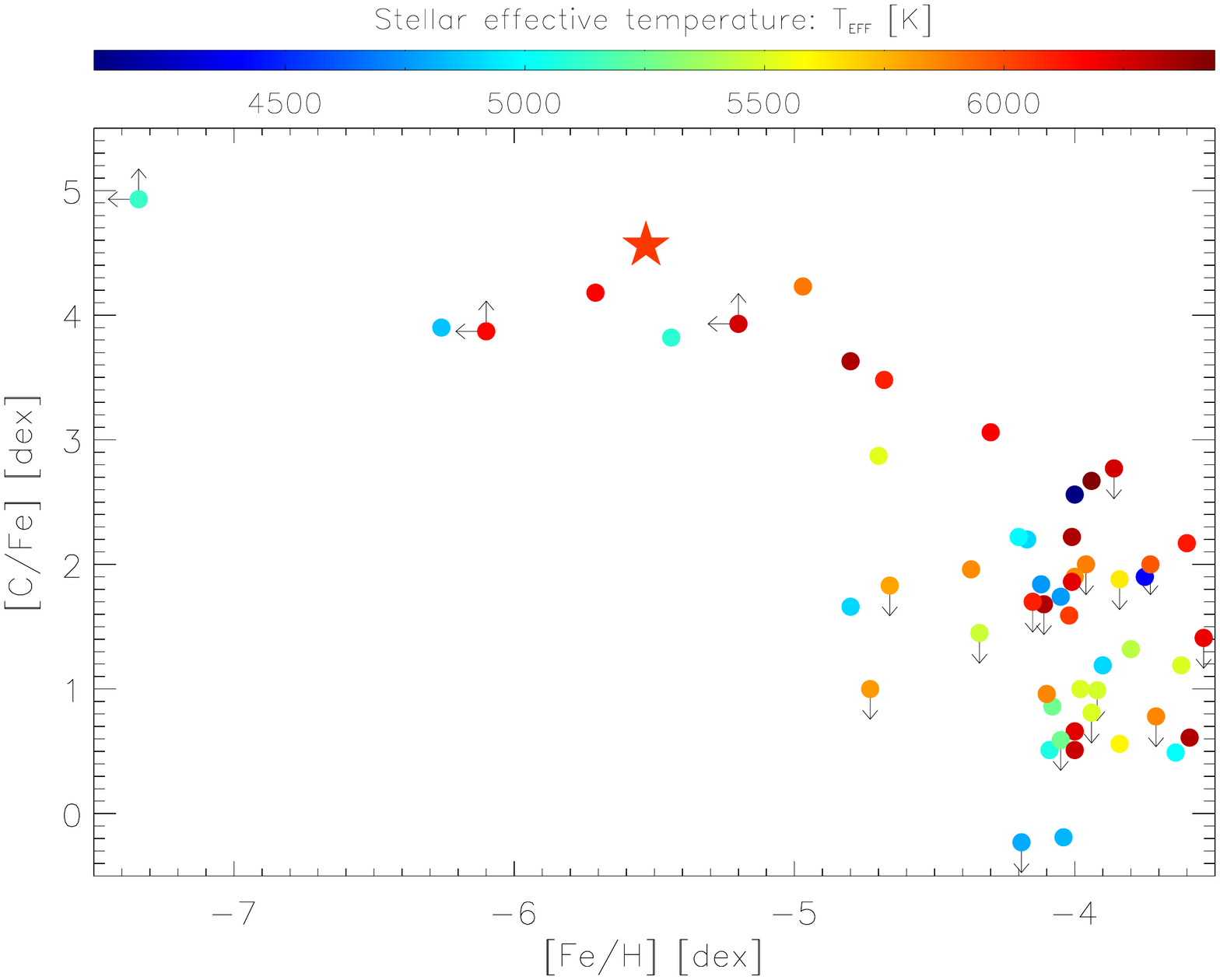}}
\end{center}
\caption{
{\it Top}: 1D-LTE oxygen-to-iron abundance ratios [O/Fe] versus metallicity [Fe/H] 
of J0815+4729 (large star symbol) compared literature measurements from the 
[\ion{O}{1}] forbidden line (diamonds), the near-IR \ion{O}{1} triplet (circles) 
and the near-UV OH lines (squares). The two triangles at [Fe/H]~$\sim -3.6$ correspond 
to the oxygen measurement from OH lines in the metal-poor binary stars 
CS 22876-032 AB~\citep{gon08aa}.
{\it Bottom}: 1D-LTE carbon abundance versus [Fe/H] of J0815+4729 (large star symbol) 
compared literature measurements for stars with [Fe/H]$< -3.5$ (small circles). 
In both panels, symbols are colored by their corresponding effective temperature.
Upward and left-pointing arrows indicate lower and upper limits. 
\label{fig:abu}
}
\end{figure}
A recent estimate of the stellar parameters 
that relies on {\it Gaia} DR2 data provides $T_{\rm eff}=6034\pm 56$~K 
and $\logg=4.6\pm0.1$~dex~\citep{ses19mnras}, consistent with our HIRES analysis.
From the accurate {\it Gaia} parallax ($\omega = 0.444\pm0.084$~mas), these authors 
derived a distance of $D=1.59\pm0.07$~kpc and
inferred an inner-halo Galactic 
orbit~\citep[see supplementary online material in][]{ses19mnras} similar 
to other iron-poor dwarf stars such as 
J0023+0307. While the highly eccentric orbit of J0023+0307 ($\epsilon = 0.88\pm0.04$) 
remains relatively close to the Galactic plane ($\left|Z\right| < 2.3$~kpc), the less 
eccentric orbit of J0815+4729 ($\epsilon = 0.32\pm0.04$) is roughly spherical with a 
radius of about 10~kpc.
We adopt $T_{\rm eff}=6050\pm100$~K and $\logg=4.6\pm0.2$~dex for our chemical 
abundance modeling.

\section{Chemical analysis} \label{sec:abu}

Chemical abundance analysis has been carried out in 1D local thermodynamic 
equilibrium (LTE) using the 
{\sc SYNPLE}\footnote{Available at \url{https://github.com/callendeprieto/synple}} 
code, and an ATLAS9 model 
atmosphere $T_{\rm eff}/\logg/$[Fe/H]~$=6050/4.6/-5$. 
We adopt a microturbulence of $\xi=1.5$~\kms\ suitable for
metal-poor dwarf stars~\citep{bar05aa}. 
We performed a fit to different spectral features of the HIRES spectrum 
(see Fig.~\ref{fig:spe}) using an automated fitting tool based on the IDL 
{\sc MPFIT}\footnote{Available at \url{http://purl.com/net/mpfit}} 
routine, with continuum location, global shift, abundance and 
global FHWM as free parameters. 
We fit the CH lines of the G-band 
in the spectral range $\lambda\lambda 426-432$~nm to obtain an average 
global (instrumental plus macroturbulent) Gaussian broadening (with no
rotation) of $v_{\rm BR}=7.5\pm0.1$~\kms, and a carbon abundance A(C)$=7.43\pm0.17$~dex. 
In Table~\ref{tab:abu} we provide the final abundances together with the 
total abundance error as the square root of the sum of 
all individual uncertainties in quadrature from $\delta T_{\rm eff}=100$~K,  
$\delta \logg=0.20$~dex, $\delta \xi =0.5$~\kms, and the statistical abundance error.
This carbon abundance is consistent
with that reported in \citet{agu18apjlI} of A(C)$\sim 7.7$~dex, obtained from 
GTC/OSIRIS medium-resolution spectroscopy, taking into account the
different effective temperature adopted in that work. We have 
also analyzed the C$_2$ feature at 516.5~nm (see Fig.~\ref{fig:oxy}), 
which shows a significantly higher
abundance, A(C)$=8.04\pm0.08$~dex. This discrepant C abundances from CH and 
C$_2$ features have been seen before in other C-enhanced metal-poor stars 
as the iron-poor star HE 0107􏰙-5240~\citep{chr04apj}.
For the rest of the elements we fixed $v_{\rm BR}=7.5$~\kms and derived 
the abundances by fitting short wavelength ranges (see Fig.~\ref{fig:oxy}). 

Table~\ref{tab:abu} summarizes our chemical abundance analysis of 
the HIRES spectrum of J0815+4729. We detect and provide 1D-LTE 
abundances for CH, C$_2$, CN, O, 
Na, Mg, Ca and Fe; and upper-limits for Li, Al, Si, Ti, Cr, Ni, Sr and Ba. 

\subsection{Iron, oxygen and nitrogen}

The iron abundance of J0815+4729 was measured from the three strongest Fe features, 
\ion{Fe}{1}~371.9~nm, \ion{Fe}{1}~382.0~nm, and 
\ion{Fe}{1}~385.9~nm (see Fig.~\ref{fig:oxy}). 
The line at \ion{Fe}{1}~371.9~nm is on the wing of a Balmer 
line that is taken into account in the synthetic spectra. The quality of the blue 
part of the spectrum is limited, with a S/N~$\sim 30$ inferred from the residuals of 
the fit. 
We thus fit the three features simultaneously to obtain a best-fit Fe abundance 
at A(Fe)~$=2.01{\pm}0.14$ in 1D-LTE. 
This Fe abundance determination makes J0815+4729 the fifth most iron-poor 
star known.  

The oxygen abundance was measured from the simultaneous fit to the three lines 
of the \ion{O}{1}~triplet at $\lambda 777$~nm (see Fig.~\ref{fig:oxy}), providing
A(O)~$=7.23{\pm}0.14$~dex.
The residuals of the fit indicate a S/N~$\sim 120$. 
We also show for comparison the UVES spectrum of HE~1327-2326~\citep{fre08apj}, where 
there is no clear detection of \ion{O}{1}~despite the higher quality (S/N$\sim250$) 
of the UVES spectrum.
The oxygen abundance of J0815+4729 is only $\sim -1.5$~dex below that of the Sun, 
indicating significant oxygen enrichment.

The nitrogen abundance was measured from the CN band at $\lambda 388.3$~nm 
(see Fig.~\ref{fig:oxy}). 
We fix the C abundance to its best-fit value from the G-band, leaving as a free 
parameter only the N abundance. The amount of nitrogen, [N/H]~$=-1.08$, in J0815+4729 
is also very high and very similar to that of the star HE~1327-2326.

\subsection{Other elements}

Calcium and sodium were measured from the resonance lines, 
\ion{Ca}{2}~K~$\lambda 393.3$~nm and \ion{Ca}{2}~H~$\lambda 396.8$~nm, and 
\ion{Na}{1}~$\lambda 588.9-589.5$. 
Magnesium abundance was derived from six lines of the two triplets at
\ion{Mg}{1}~$\lambda 382.9-383.8$~nm and \ion{Mg}{1}~$\lambda 516.7-518.3$~nm. 
The typical dispersion of individual line abundances for Ca and Mg is 0.1~dex and 
0.05~dex for Na, whereas the statistical error from individual fits is 0.04~dex.
For the rest of the elements, we were only able to obtain upper-limits from
the strongest lines typically present in the spectra of iron-poor dwarf stars. 
In Table~\ref{tab:abu} we provide the abundance upper-limit estimates together with 
the wavelength of the line used for each element. 

\subsection{Lithium}

We were unable to detect lithium in the HIRES spectrum of J0815+4729. We derived an
upper-limit of A(Li)~$< 1.3$~dex (see Fig.~\ref{fig:oxy}) from the lithium doublet 
\ion{Li}{1}~$\lambda 670.8$~nm. Our J0815+4729 spectrum with a S/N$\sim 120$ is 
compared with the spectrum of HE~1327-2326, which has a lithium abundance limit 
A(Li)~$< 0.7$~dex~\citep{fre08apj}, in contrast to the unambiguous lithium detection 
and abundance measurement of A(Li)~$=2.0$~dex found in the unevolved star J0023+0307 at 
the lowest iron abundance ([Fe/H]~$<-6.1$, \cite{agu19apjl}).

\section{Discussion and conclusions} \label{sec:dis}

\begin{figure}
\begin{center}
{\includegraphics[clip=true,width=85mm,angle=0]{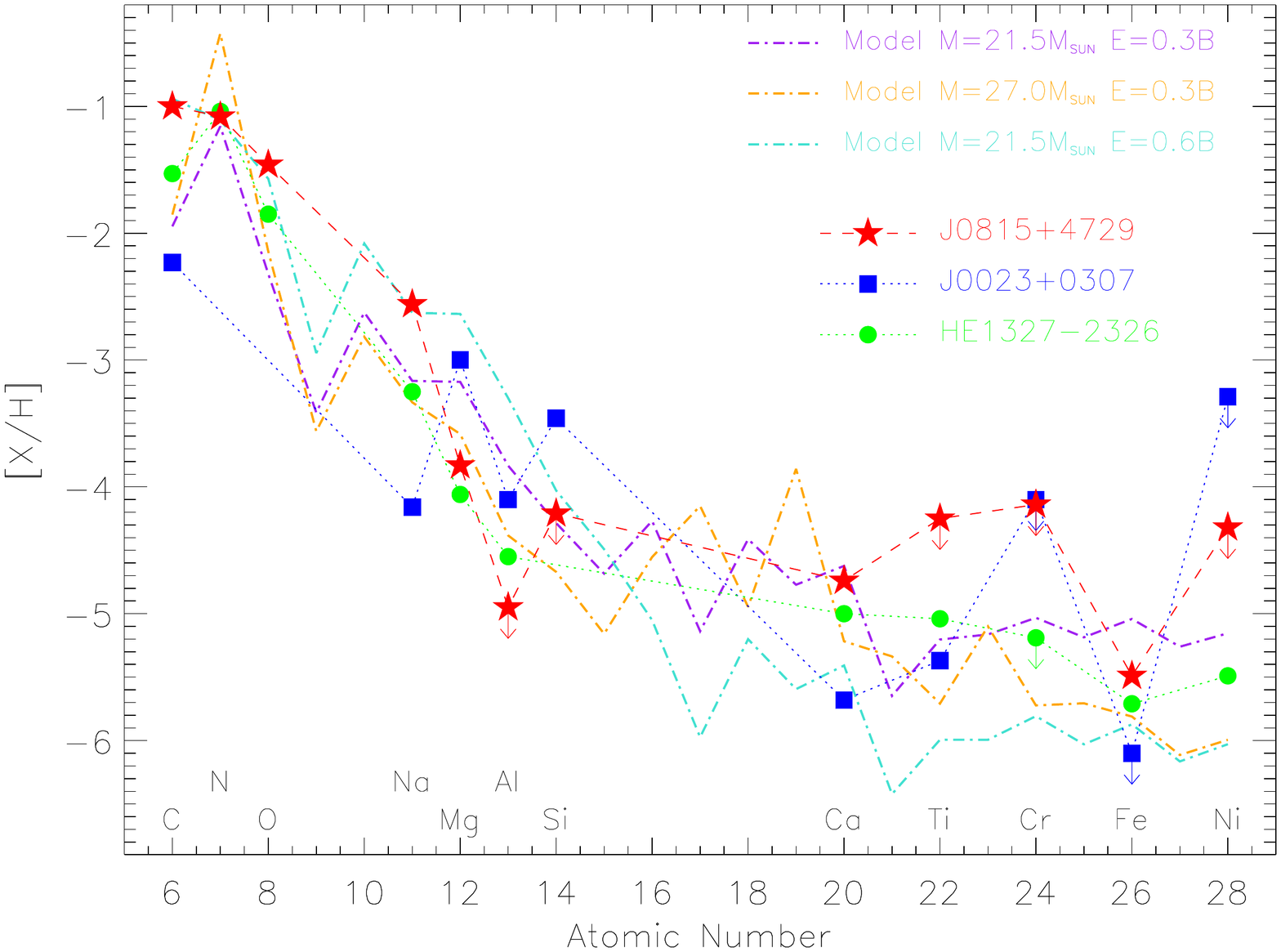}}
{\includegraphics[clip=true,width=85mm,angle=0]{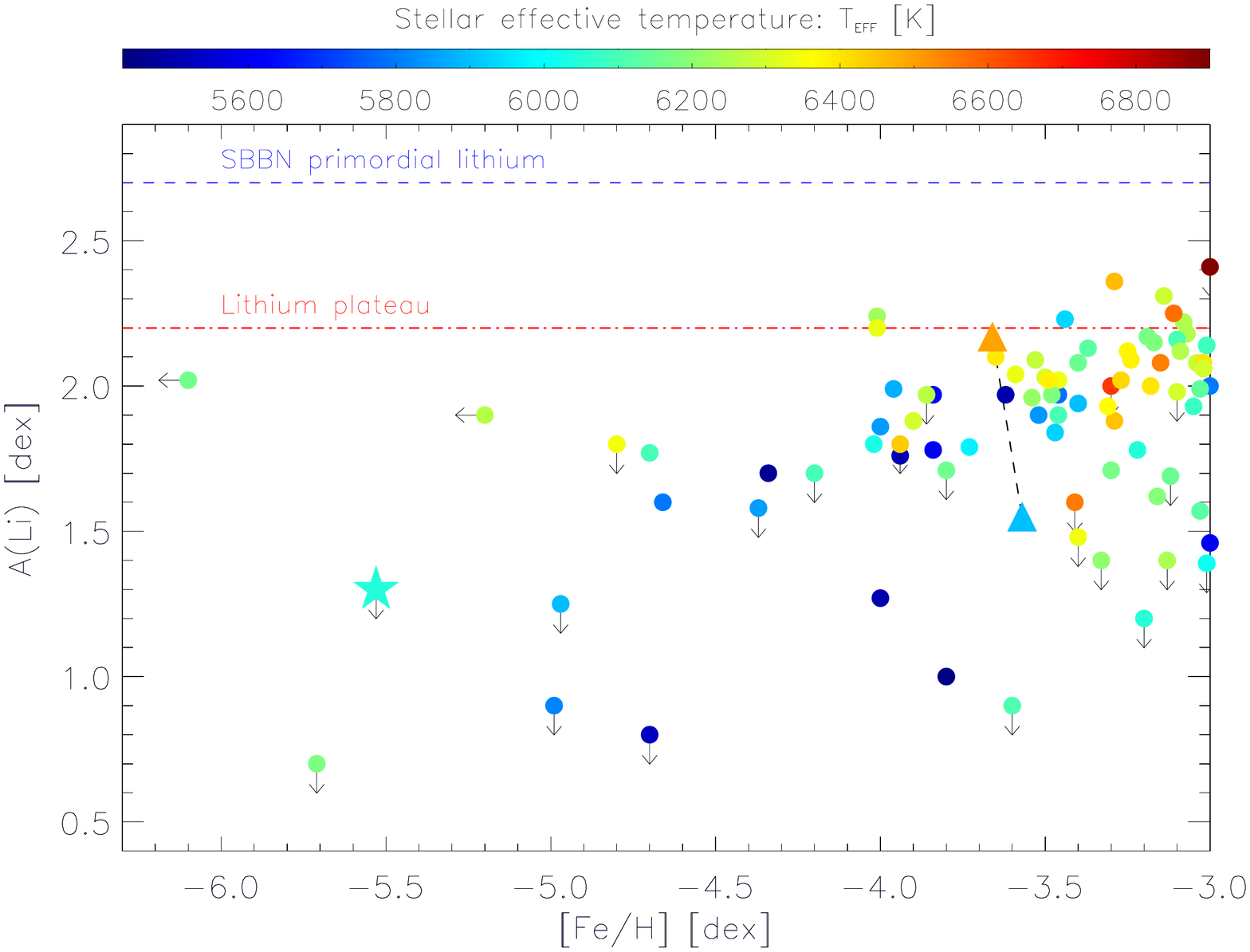}}
\end{center}
\caption{
{\it Top}: 1D-LTE abundance pattern of J0815+4729 (red stars) compared to the abundance 
pattern of two iron-poor stars with similar effective temperature and iron content: 
J0023+0307~\citep[blue squares]{agu19apjl} and HE~1327-2326~\citep[green 
circles]{fre08apj}. Upper limits are indicated by downward arrows.
{\it Bottom}: lithium abundance versus [Fe/H] for 
J0815+4729 compared with literature measurements for unevolved stars with [Fe/H]$<-3.0$ 
and logg$>3.0$. 
In both panels, symbols are colored by their corresponding effective temperature.
The two large triangles connected with a black dashed line correspond 
to the metal-poor binary CS 22876-032 AB \citep{gon08aa,gon19aa}.
Downward and left-pointing arrows indicate upper limits. 
The blue dashed line is the predicted SBBN 
primordial lithium abundance and the red dash-dotted line is the 
value of the lithium plateau, known as the  {\it Spite Plateau}.
\label{fig:ali}
}
\end{figure}

The detection of the oxygen triplet in the spectrum of J0815+4729 is the 
first such detection in an extremely iron-poor star, anticipating a significant 
oxygen enhancement in this star with an oxygen-to-iron ratio [O/Fe]~$=4.03\pm0.12$.
The lowest metallicity for which the \ion{O}{1}~had been previously detected 
is at [Fe/H]~$\sim -3.8$ for the metal-poor giants CS 22949-037 and CS 29498-043,
with oxygen abundances A(O)~$=8.3$~dex~\citep{isr04aa}; and at [Fe/H]~$=-3.9$ 
for the metal-poor dwarf SDSS~J2209-0028~\citep{spi13aa}
with an oxygen abundance A(O)~$=7.0$~dex and [O/Fe]~$\sim2.2$.

In Fig.~\ref{fig:abu} we display the 1D-LTE oxygen-to-iron ratio, [O/Fe], versus the
metallicity, [Fe/H], of J0815+4729 compared with literature measurements of other
stars. This figure includes only 1D-LTE oxygen measurements from different oxygen 
indicators, the [\ion{O}{1}] forbidden line, the near-IR \ion{O}{1} triplet, 
and the near-UV OH lines. 
A trend of increasing [O/Fe] towards lower metallicities is apparently seen. 
The [O/Fe] trend versus metallicity has been a matter of long discussions in the 
literature also related with the disagreement between different abundance
indicators~\citep[see e.g.][and references therein]{gon10aa}. 
Recent results in 3D and NLTE seem to confirm an increasing trend 
[O/Fe] towards lower metallicities, in particular, at the lowest
metallicities for all O abundance indicators\citep{gon10aa,ama15mnras}.
The upturn of [O/Fe] at the lowest metallicities ($-3.5<$[Fe/H]$<-2.5$) may 
indicate a shift towards more massive supernovae (SNe), more metal-poor SNe and 
hypernovae (HNe) at earlier epochs, thus expected to yield larger [O/Fe] 
ratios~\citep{kob06apj}.

The two giant stars CS 22949-037 and CS 29498-043 also show high Mg and Si abundances, 
and lower Na and Al abundances, 
displaying an ``odd-even'' pattern also seen in the metal-poor stars 
J2217+2104~\citep{aok18pasj} and J0023+0307~\citep{agu19apjl}.
This odd-even pattern has been associated with massive 
pair-stability supernovae~\citep[PISNe, e.g.][]{heg02apj}. 
Fig.~\ref{fig:ali} compares the abundance pattern of J0815+4729 to those 
of the two iron-poor stars HE~1327-2326 and J0023+0307, which have similar stellar 
parameters and iron content. 
PISNe models predict element abundance ratios [Na/Mg]~$=-1.5$ and 
[Ca/Mg]~$\sim 0.5-1.3$~\citep{tak18apj}. While one each of these ratios, 
[Na/Mg]$=-1.16$ in J0023+0307 and [Ca/Mg]$=0.94$ in HE~1327-2326, are marginally 
consistent with PISNe predictions, the observed abundance ratios of J0815+4729, 
[Na/Mg]~$=1.27$ and [Ca/Mg]~$=-0.91$ rule out any contamination from PISNe.

J0815+4729 shows a particularly large enhancement of CNO elements with 1D-LTE ratios 
[X/Fe] of 4.49, 4.41 and 4.03 for C, N and O respectively. 
J0815+4729 shows significantly higher C abundance than HE~1327-2326 and J0023+0307 
(see Figs.~\ref{fig:abu} and~\ref{fig:ali}). 
The amount of N is slightly lower than C in J0815+4729 while HE~1327-2326 shows 
significantly higher N to C abundance by 0.5~dex. 
Oxygen, however, is lower than C by 0.5 and 0.3~dex in J0815+4729 and 
HE~1327-2326 respectively.
There is so far no evidence for RV variations nor any chemical signature of mass 
transfer from (prior) companion AGB stars to any of these three metal-poor dwarfs.
Their location in the A(C) vs [Fe/H] diagram is coincident with CEMP-no~\citep{bon15aa} 
stars, which are predicted to be the direct descendants of the first generation of stars, 
with atmospheric abundances (including vast amounts of carbon) reflecting 
nucleosynthesic yields from one or a few zero-metallicity core collapse SNe diluted 
with the primordial gas. 
J0815+4729 appear to be consistent with the upper-envelope of the low-carbon 
band in the A(C)-[Fe/H] diagram~\citep{spi13aa,yoo16apj}. 
The extremely high CNO abundances, coupled with 
high Na and Mg abundances (see Fig.~\ref{fig:ali}), suggests a fallback model in which 
some of the Ca and Fe fall back into the black hole~\citep{ume03nat}. The low 
[Mg/C]~$=-2.83$ and high [Mg/Fe]~$=1.66$ and [Ca/Fe]~$=0.75$ ratios in J0815+4729 can 
also be explained by a mono-enriched scenario (i.e. one SN per mini-halo) based on the 
divergence model of the chemical displacement (DCD) from \citet{har18mnras}. 
On the other hand, the [Al/Mg]~$<-1.12$ and [N/Na]~$=1.48$ abundance ratios suggest 
a multi-enriched scenario.
In either case, the similarity between the abundance patterns of J0815+4729 and 
HE~1327-2326 points to enrichment from a Pop~III SN progenitor with mass between 
15~\msun~\citep{ish18apj} and 27~\msun~\citep{heg10apj}.
We have used the {\sc StarFit}\footnote{http://starfit.org} code~\citep{heg10apj} 
to fit the observed abundances.
In Fig.~\ref{fig:ali} we show several metal-free SN models with masses 21.5-27~\msun and
energies 0.3-0.6B (1~B = 1~Bethe = 10$^{51}$~erg). All these three low-energy SN models 
with very little mixing seem to fit reasonably well the observed abundances.

Metal-poor damped Lyman $\alpha$ (DLA) systems show abundance ratios [C/O]~$\sim -0.3$,
well explained with core-collapse Pop~III SN models (with low kinetic energies 
at $1\times10^{51}$~erg) and a typical progenitor mass of 20~\msun~\citep{coo17mnras}. 
Similarly, halo metal-poor dwarf stars~\citep[e.g.][]{ama19aaoc} have 
[C/O]~$\sim -0.5$, consistent with  DLAs.
The 1D-LTE [C/O] abundance ratio of J0815+4729 is $\sim 0.5$, and thus significantly 
higher than both ratios, but this value may be affected by 3D-NLTE effects that would 
possibly lower this ratio~\citep{beh10aa,ama19aaoc}.

The absence of detectable lithium in the spectrum of J0815+4729 at a limit of 
A(Li)~$< 1.3$~dex exacerbates the cosmological lithium depletion problem. 
Fig.~\ref{fig:ali} displays the Li abundances of unevolved Galactic stars with 
[Fe/H]~$< -3$. Most of these stars have 
A(Li)~$\sim 2.0$~dex, with different degree of Li depletion at increasingly higher
temperature and decreasing metallicity~\citep{bon18aa,gon19aa}.
There is a clear lack of stars between this upper envelope defined by the 
lithium plateau~\citep[A(Li)~$\sim 2.2$;][]{spi82nat,reb88aa} 
and the primordial Li abundance predicted by the standard Big Bang nucleosynthesis 
(SBBN)~\citep[A(Li)~$\sim 2.7$;][]{cyb16rmp}. 

At [Fe/H]~$< -3.5$, only three stars show Li measurements at 
A(Li)~$=2.2$~\citep{gon08aa,gon19aa,bon18aa}.
\citet{mat17pasj} have suggested that most stars at [Fe/H]~$<-4.5$ show Li
abundances or upper-limits A(Li)~$<1.8$~dex, and conclude that there is no clear 
connection between the amount of carbon and the Li content. 
Among the four unevolved stars with [Fe/H]~$<-5$, all C-enhanced, only two stars have 
Li detections with abundances close to the Li plateau: J0023+0307 with 
A(Li)~$\sim 2.0$~dex~\citep{agu19apjl} and J0135+0641 with 
A(Li)~$\sim 1.9$~dex~\citep{bon18aa}.  HE~1327-2326 and, J0815+4729 with the strongest 
carbon enhancement among stars with [Fe/H]~$<-5$, do not show any Li detection. 
Any model claiming a solution to the cosmological 
lithium problem must explain the more than 1~dex Li abundance
variance among these stars.

\acknowledgments
We thank Observing Assistant Heather Hershley and Support Astronomer Sherry Yeh 
at Keck Observatory for their assistance with the observations. 
JIGH acknowledges financial support from the Spanish Ministry of Science, 
Innovation and Universities (MICIU) under the 2013 Ram\'on y Cajal program 
RYC-2013-14875. JIGH, CAP and RR acknowledge financial support 
from the Spanish Ministry project MICIU AYA2017-86389-P. 
DA thanks the Leverhulme Trust for financial support. 

%




\bibliography{mpoor}

\end{document}